\documentclass[a4paper,12pt]{article}
\usepackage{pos}
\usepackage{xspace}
\usepackage{booktabs} 
\newcommand{\Qsq}{\ensuremath{Q^2}\xspace}
\newcommand{\MeV}{\ensuremath{\mathrm{MeV}}\xspace}
\newcommand{\GeV}{\ensuremath{\mathrm{GeV}}\xspace}
\newcommand{\TeV}{\ensuremath{\mathrm{TeV}}\xspace}
\newcommand{\GeVsq}{\ensuremath{\mathrm{GeV}^2}\xspace}

\newcommand{\sw}{\ensuremath{\sin^2\hspace*{-0.15em}\theta_W}}
\newcommand{\sweff}{\ensuremath{\sin^2\hspace*{-0.15em}\theta_{\textrm{W},f}^\textrm{eff}}} 
\newcommand{\sweffl}{\ensuremath{\sin^2\hspace*{-0.15em}\theta_{\textrm{W},\ell}^\textrm{eff}}} 

\newcommand{\gf}{\ensuremath{G_{\rm F}}}

\newcommand{\ad} {\ensuremath{g_A^d}}
\newcommand{\vd} {\ensuremath{g_V^d}}
\newcommand{\au} {\ensuremath{g_A^u}}
\newcommand{\vu} {\ensuremath{g_V^u}}
\newcommand{\aq} {\ensuremath{g_A^q}}
\newcommand{\vq} {\ensuremath{g_V^q}}
\newcommand{\gae}{\ensuremath{g_A^e}}
\newcommand{\ve} {\ensuremath{g_V^e}}

\newcommand{\mw}{\ensuremath{m_W}}
\newcommand{\mW}{\mw}
\newcommand{\mz}{\ensuremath{m_Z}}
\newcommand{\mZ}{\mz}
\newcommand{\mt}{\ensuremath{m_t}}

\usepackage{parskip} 

\begin{document}

\title{Precision electroweak physics at the LHeC and FCC-eh}

\author[a]{Daniel Britzger}
\author[b]{Max Klein}
\author[c]{Hubert Spiesberger\vspace{0.4cm}
}
\affiliation[a]{Max-Planck-Institut für Physik, Föhringer Ring 6, 80805 München, Germany}
\affiliation[b]{University of Liverpool, Oxford Street, UK-L69 7ZE Liverpool, United Kingdom}
\affiliation[c]{Institut f{\"u}r Physik, Johannes-Gutenberg-Universit{\"a}t, Staudinger Weg 7, 55099 Mainz, Germany}



\abstract{
The proposed electron-proton collider experiments LHeC and FCC-eh at CERN
are the highest resolution microscopes that can be realised in the
present century and they would represent a really unique research facility.
We exploit simulated neutral-current and charged-current deep-inelastic
scattering data of the LHeC and the FCC-eh and examine their
sensitivity for precision physics in the Electroweak sector of the
Standard Model (SM), like the effective weak mixing angle \sweffl, or the
light-quark weak-neutral-current couplings.
Unique measurements are further feasible at high precision for the running
of the weak mixing angle, as well as for electroweak effects in
charged current interactions.
The sensitivity to beyond SM effects is studied using the generic $S$, $T$
and $U$ parameterization.
The report summarizes previous studies about the LHeC and presents
new prospects for the FCC-eh.
}

\FullConference{%
  Proceedings submitted to the EPS Conference on High Energy Physics (EPS-HEP2021),\\
  Hamburg, Germany, 26-30 July 2021
}


\maketitle

\section{Introduction}
\noindent
The LHeC~\cite{AbelleiraFernandez:2012cc,LHeC:2020van} and 
FCC-eh~\cite{Abada:2019lih} are proposed electron-proton colliders at CERN
with center-of-mass energies of 1.3\,\TeV or 3.5\,\TeV, respectively.
Unlike HERA, or other previous proposals, they are designed as
high-luminosity colliders with a targeted integrated luminosity that
exceeds $\mathcal{L}>1\,\text{ab}^{-1}$, and thus provides great new
opportunities for precision measurements in deep-inelastic
electron-proton scattering at high energy scales.

In this report, we study simulated inclusive neutral- and
charged-current deep-inelastic scattering (NC and CC DIS) data in
electron-proton collisions with respect to their sensitivity to parameters of the electroweak interaction in the Standard Model (SM) or possible generic extensions.
The process of NC and CC DIS are mediated through spacelike momentum
transfer and therefore such measurements are complementary to measurements in the
timelike domain in $e^+e^-$ or $pp$ collisions.

After a brief recapitulation of the formalism, we present new prospects for the
FCC-eh and compare them with LHeC expectations from
Refs.~\cite{Britzger:2020kgg,LHeC:2020van}.

\section{Electroweak physics in DIS}
At the LHeC and FCC-eh, inclusive NC and
CC DIS cross sections will be measured in $ep$ collisions as a function of \Qsq and $x$,
where \Qsq\ is the virtuality of the exchanged boson and $x$ is the
Bjorken scaling variable.
These data will be compared to predictions in the framework of the
SM.
In the following, we will briefly sketch the calculation of the NC DIS
cross sections, and where higher-order Electroweak (EW) or QCD corrections
contribute. For details of CC scattering, we refer to Ref.~\cite{Britzger:2020kgg}.

Inclusive NC DIS cross sections are expressed in terms of
generalized structure functions $\tilde{F}_2^\pm$,
$x\tilde{F}_3^\pm$ and $\tilde{F}_{\rm L}^\pm$ at EW
 leading order (LO) as
\begin{equation}
\frac{d^2\sigma^{\rm NC}(e^\pm p)}{dxd\Qsq}
=
\frac{2\pi\alpha^2}{xQ^4}
\left[Y_+\tilde{F}_2^\pm(x,\Qsq)
     \mp Y_{-} x\tilde{F}_3^\pm(x,\Qsq)
     - y^2 \tilde{F}_{\rm L}^\pm(x,\Qsq)
\right]~,
\label{eq:cs}
\end{equation}
using the fine structure constant  $\alpha$, the Bjorken scaling
variable $x$, the inelasticity $y$, and the factors $Y_\pm =
1\pm(1-y)^2$ express the helicity dependence of the process.
The generalized structure functions represent the contributions
from pure $\gamma$- and $Z$-exchange, and its
interference~\cite{Klein:1983vs} like 
\begin{eqnarray}
  \tilde{F}_2^\pm
  &=& F_2
  -(\ve\pm P_e\gae)\varkappa_ZF_2^{\gamma Z}
  +\left[(\ve\ve+\gae\gae)\pm2P_e\ve\gae\right]\varkappa_Z^2F_2^Z~,
\label{eq:strfun1}
  \\
  \tilde{F}_3^\pm
  &=& ~~~~
  -(\gae\pm P_e\ve)\varkappa_ZF_3^{\gamma Z}
  +\left[2\ve\gae\pm P_e(\ve\ve+\gae\gae)\right]\varkappa_Z^2F_3^Z~,
\label{eq:strfun2}
\end{eqnarray}
where $P_e$ is the longitudinal polarization of the electron beam.  
A similar decomposition exists for $\tilde{F}_L$.
The coefficient $\varkappa_Z$ accounts for the $Z$-boson
propagator and the normalization of the weak relative to the
electromagnetic interaction.
In the on-shell renormalization scheme it is calculated from the weak
boson masses \mz\ and \mw\  at LO as 
\begin{equation}
  \varkappa_Z(\Qsq)
  = \frac{\Qsq}{\Qsq+m^2_Z}
  \frac{1}{4\sw \cos^2\theta_W}\,,
\label{eq:kappaZ-LO}
\end{equation}
using $\sw = 1 - \cos^2\theta_W = 1 - m_W^2/m_Z^2$.
Higher-order corrections are included through an additional factor
proportional to $1/(1-\Delta r)$~\cite{Sirlin:1980nh}. 
In the on-shell scheme also the Fermi constant \gf\ is a prediction. 
Electroweak couplings at the hadronic vertex are included in the structure
functions and one can write them, in the naive quark-parton
model, in terms of quark and anti-quark
parton distribution functions, $q(x)$ and $\bar{q}(x)$,
\begin{eqnarray}
  \left[F_2,F_2^{\gamma Z},F_2^Z\right]
  &=&
  x\sum_q\left[Q_q^2,2Q_q\vq,\vq\vq+\aq\aq \right]\{q+\bar{q}\}~,
\label{eq:last1}
  \\
  x\left[F_3^{\gamma Z},F_3^Z\right]
  &=&
  x\sum_q\left[2Q_q\aq,2\vq\aq\right]\{q-\bar{q}\}\,,
\label{eq:last2}
\end{eqnarray}
where $Q_f$ denotes the electric charge of a fermion $f$ (with $f = e$,
$q$ and $q = u$, $d$). 
The vector and axial-vector couplings of a fermion to the $Z$-boson,
$g_V^f$ and $g_A^f$,  are given by
\begin{eqnarray}
  g_A^{f}
  &=& \sqrt{\rho_{\text{NC}, f}} \, I^3_{{\rm L},f}
  \label{eq:gA-NLO} \,, \\
  g_V^{f}
  &=& \sqrt{\rho_{\text{NC}, f}} \, \left(I^3_{{\rm L},f} - 2
  Q_{f} \, \kappa_{f} \, \sw \right)\,,
\label{eq:gV-NLO}
\end{eqnarray}
where $I^3_{{\rm L},f}$ is the third component of the weak-isospin
of the fermion $f$.
The form factors $\rho_{\text{NC}, f}$ and $\kappa_{f}$ are unequal to
unity beyond the LO and include higher-order corrections due to the
$\gamma Z$ mixing and $Z$ self-energy corrections~\cite{Bohm:1986na,Bardin:1988by,Hollik:1992bz}.
They are scale dependent and also include flavor dependent contributions
beyond the universal correction factors.

Subject of interest in EW physics are the Born-level quantities, as
well as potential modifications to the higher-order virtual
corrections, which could originate from new physics.

\section{Methodology}
For our purpose of studying the sensitivity of future inclusive NC
and CC DIS measurements to EW parameters, we simulate
double-differential cross section measurements including a full set of
statistical and systematic uncertainties.
The procedure employs a numerical
simulation procedure~\cite{Blumlein:1990dj} and follows closely the one for the
LHeC~\cite{Britzger:2020kgg,LHeC:2020van}.
A full suite of all cross section measurements is important, since
also the PDFs are determined simultaneously from these data.
The simulated data are summarized in Tab.~\ref{tab:data}.

NC and CC DIS cross section measurements are simulated
in $e^+p$ and $e^-p$ operation mode and different
lepton-beam polarization states
for the nominal FCC-hh proton-beam energy
of $E_p=50\,\TeV$.
A data set with reduced integrated luminosity is simulated
for $E_p=7\,\TeV$, which will be used for $F_L$ measurements.
The kinematic range is assumed to be $\Qsq\ge3.5\,\GeVsq$ in NC, and
$\Qsq\geq50\,\GeVsq$ in CC.
Altogether 1117 cross section values are simulated for NC
and CC DIS.
Due to computational reasons, we have merged always 2$\times$2 data
points, so that these are representative for altogether 4468 cross
section measurements.

\noindent
\begin{minipage}[t]{.42\textwidth}
  \centering
  \footnotesize
  \begin{tabular}[t]{lccccc}
    %
    %
    %
    %
    \toprule
    $E_p$ [TeV] & $P_e$ & Charge & $\mathcal{L}$\,[ab$^{-1}$] \\
    \midrule
    50 & -0.8 & -1 & 2 \\
    50 & +0.8 & -1 & 0.5 \\
    50 &  0   & +1 & 0.2 \\
     7 & -0.8 & -1 & 0.2 \\
    \bottomrule
 \end{tabular}
  \captionof{table}{
    Simulated data sets for different proton beam energies $E_p$,
    electron beam polarization $P_e$, electron or positron beam, and prospected
    integrated luminosity $\mathcal{L}$.
    All data will become available  for lepton beam energies of $E_e=60\,\GeV$ and for NC and CC DIS.
}
\label{tab:data}
\end{minipage}
\hfill\begin{minipage}[t]{.54\linewidth}
  \centering
  \footnotesize
  \begin{tabular}[t]{lc}
    \toprule
    Source of uncertainty & Uncertainty \\
    \midrule
    Scattered electron energy scale $\Delta E_e' /E_e'$ & 0.1 \% \\
    Scattered electron polar angle  & 0.1\,mrad \\
    Hadronic energy scale $\Delta E_h /E_h$ & 0.5\,\% \\
      \addlinespace
    Radiative corrections & 0.3\,\% \\
    Photoproduction background (for $y > 0.5$) & 1.0\,\% \\
    Global efficiency error & 0.5\,\%  \\
    Luminosity uncertainty & 1.0\,\% \\
    Statistical uncertainty & $\geq0.1\,\%$ \\
    \bottomrule
 \end{tabular}
  \captionof{table}{
Assumptions used in the simulation of the DIS cross sections
for the size of uncertainties from various sources. The top three are
calibration uncertainties which are transported to
provide correlated systematic cross section errors.
The lower five values are uncertainties of the cross section
caused by various sources.
}
\label{tab:sys}
\end{minipage}

The data include a set of the most important experimental
uncertainties.
These are summarised in Tab.~\ref{tab:sys}.
The statisical uncertainties are set to a minimum value of 0.1\,\%.
The hadronic energy scale uncertainty of 0.5\,\% becomes important for
CC and for NC DIS at low-$y$, where the electron reconstruction method
worsens.
Heavy-flavor cross sections are not included in our study, although
they may yield an interesting sensitivity to anomalous $Zb\bar{b}$
couplings~\cite{Yan:2021htf}.

The expected uncertainties in the electroweak parameter of interest
are determined in a combined fit together with the PDFs.
The PDFs are parameterised with five functions at a scale of
$Q_0=1.38\,\GeV$, and after the application of sum rules 13 free
parameters are associated to the PDFs and determined from the FCC-eh
pseudo-data.
The structure functions are calculated in NNLO QCD using the program
QCDNUM~\cite{Botje:2010ay}, the electroweak 1-loop corrections are
implemented in the program EPRC~\cite{Spiesberger:1995pr}, and the
cross section calculation is implemented in a private code.
The program Minuit~\cite{James:1975dr} is used for the calculation of
the Hesse matrix of the Azimov fit, and further technical details are given in
Ref.~\cite{Britzger:2020kgg}.
This kind of algorithm is termed PDF+{\em EW} fit in the following, where
{\em EW} will be replaced with the parameter(s) of interest.
As a consequence of the simultaneous determination of the PDFs and the
EW parameters, the prospected uncertainties will always include the
experimental and PDF-related uncertainties.
The latter, however, may become comparatively small, since the PDFs are
predominantly constrained from low-\Qsq\ data.
In contrast, high-\Qsq\ data are particularly sensitive to EW parameters.

The results from FCC-eh pseudo data will be compared with
expectations from LHeC~\cite{Britzger:2020kgg}.
These are available for two electron-beam energy configurations of 50
or 60\,\GeV\ and will be termed \emph{LHeC-50} and \emph{LHeC-60}, respectively.

\section{Precision electroweak physics with inclusive NC and CC DIS}
In the following results from the PDF+EW fits to the simulated FCC-eh
pseudo-data are presented.
The expected uncertainties include the experimental
uncertainties and PDF-related uncertainties.

\subsection{Weak-neutral-current couplings of the first generation quarks}
The couplings of fermions to weak-neutral-currents can generically be
considered as free parameters.
While the couplings of the leptons and heavy quarks were measured at
LEP/SLD with high precision~\cite{ALEPH:2005ab}, the couplings of the first generation quarks
have sizeable experimental uncertainties still today.
In the Standard Model, these couplings are determined as vector and
axial-vector couplings (c.f.\ eqs.~\eqref{eq:gA-NLO} and~\eqref{eq:gV-NLO}).
A meaningful test of the Standard Model is performed by determining
$g_V$ and $g_A$ and comparing them with the SM prediction.
High experimental precision is desirable, since in the SM
higher-order corrections modify the vector and axial-vector couplings,
and new physics may alter these virtual corrections.

\begin{figure}[tbh!]
    \centering
    \includegraphics[width=0.4\textwidth]{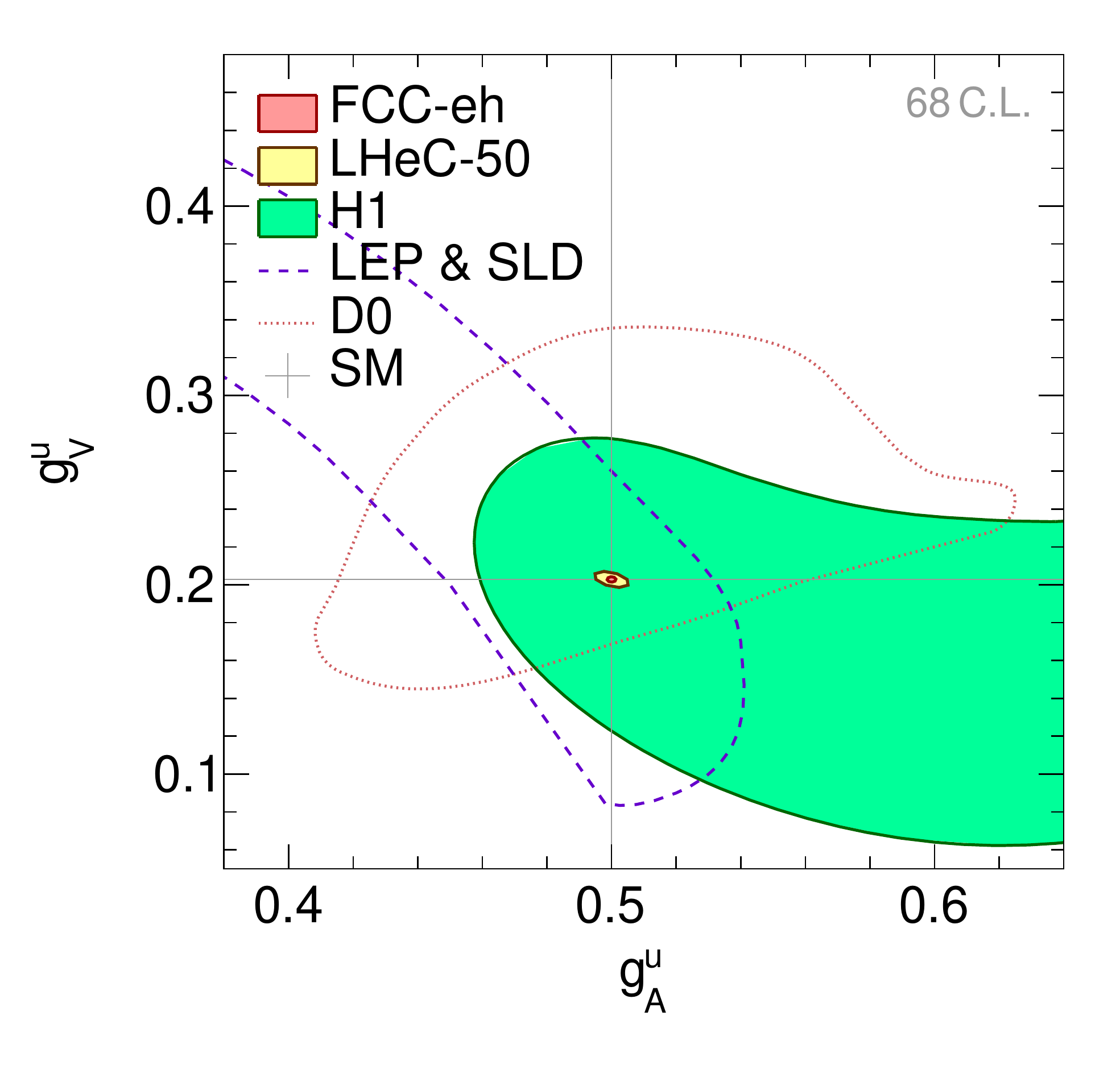}
    \includegraphics[width=0.4\textwidth]{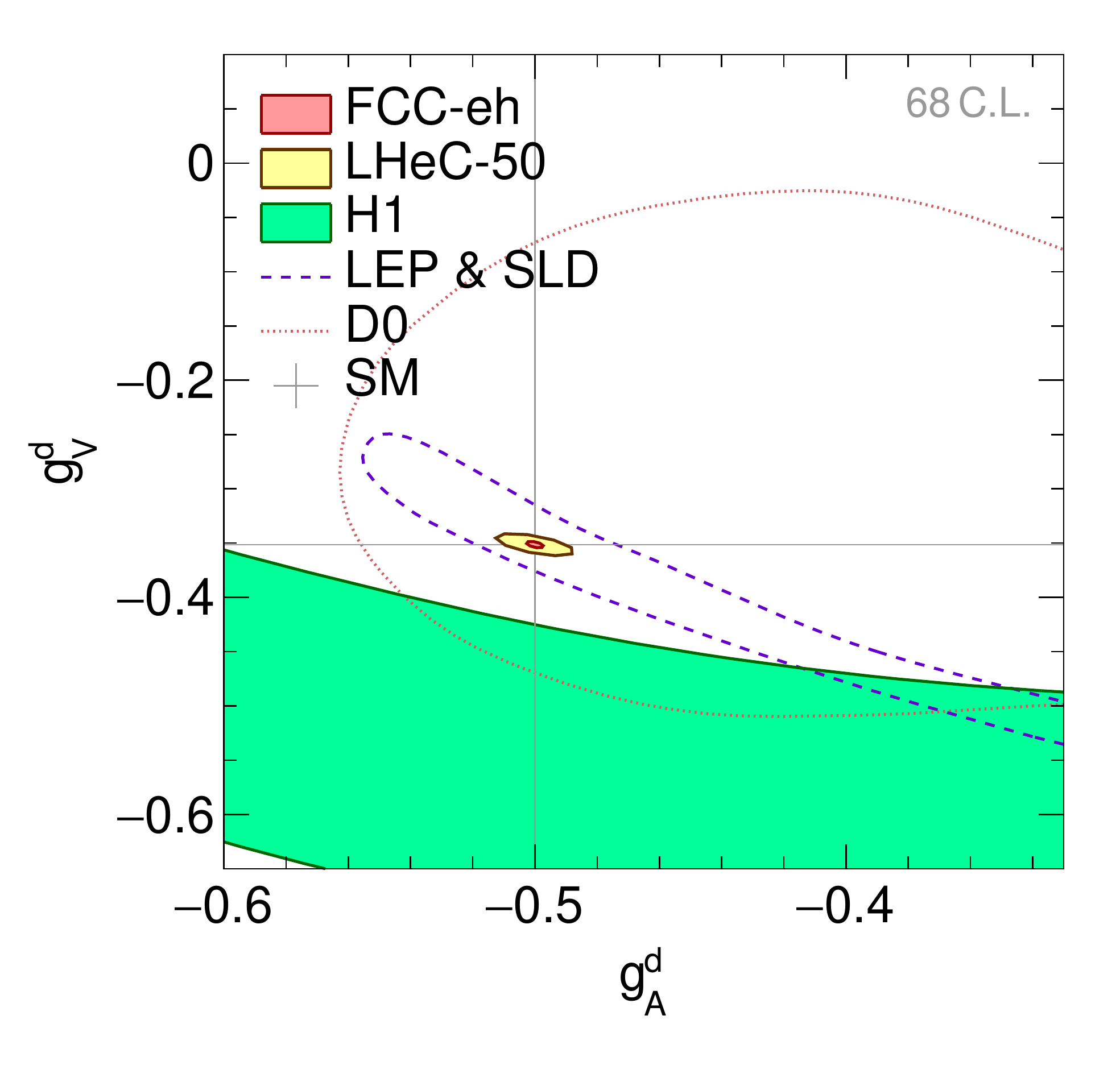}
    \caption{
      Prospected uncertainties of the weak-neutral-current vector and
      axial-vector couplings of first generation quarks  (up-type, and
      down-type) to the $Z$-boson at 68\,\% confidence level.
      Left: down-type quarks. Right: up-type quarks.
      The prospected uncertainties are compared with results from
      LEP+SLD~\cite{ALEPH:2005ab}, D0~\cite{Abazov:2011ws} and H1~\cite{Spiesberger:2018vki}.
      The SM expectations are at the crossing of the horizontal and vertical lines.
    }
    \label{fig:couplings}
\end{figure}
We perform a PDF+\ad+\vd+\au+\vu\ fit to study the determination the NC couplings of
the $u$-type and $d$-type quarks.
The contributions from strange quarks are smaller than 10\,\% and those from
heavier quarks are negligible. Thus the parameters obtained in this fit can be interpreted
as the NC couplings of the first generation quarks.
The lepton couplings are kept fixed at their predicted values since they were
tested with high precision at LEP/SLD.
The expected uncertainties from a fit to LHeC or FCC-eh pseudo-data
are presented in Figure~\ref{fig:couplings}.
The uncertainties are found to be smaller by orders of magnitudes
than today's best measurements~\cite{ALEPH:2005ab,Abazov:2011ws,Spiesberger:2018vki}.
This is due to the high luminosity and the high center-of-mass energy
of the LHeC and FCC-eh, such that weak interactions of the valence
quarks are probed.

\subsection{The effective weak mixing angle}
The weak mixing angle is the central parameter in the electroweak theory~\cite{EW:PDG2020}.
It determines the value of $g_V$, but also higher-order
corrections become relevant therein.
As a consequence, usually the \emph{effective}
weak mixing angle is studied, which combines the Born-level expression and
higher-order corrections as
\begin{equation}
  \sin^2 \theta_{W,f}^{\rm eff} (\mu^2) =
  \kappa_{f}(\mu^2) \sw \,.
\label{eq:sin2w-eff}
\end{equation}
Its value at $\mu^2=m_Z^2$ is subject to measurements at LEP/SLD, Tevatron and the LHC.
We estimate the uncertainty of \sweffl\ in a PDF+\sweffl\ fit. %
In order to map the values of \sweff\ that contribute in the spacelike
calculation of the NC DIS cross sections ($\mu^2 = -Q^2$) to the
definition of \sweffl\ from LEP/SLD, we utilize scale- and flavor-dependent
correction factors and assume the validity of the SM for that purpose.
We find, that from FCC-eh pseudo-data, the value of \sweffl\ can be
determined with an uncertainty of  
\begin{equation}
  \Delta\sweffl = \pm\,0.00011\,.
\end{equation}
This uncertainty includes experimental and PDF-related
uncertainties,  and the results are compared with presently available  measurements and the
PDG-prediction~\cite{EW:PDG2020} in Fig.~\ref{fig:mwsw2} (right).

\begin{figure}[tbh!]
    \centering
    \includegraphics[width=0.50\textwidth]{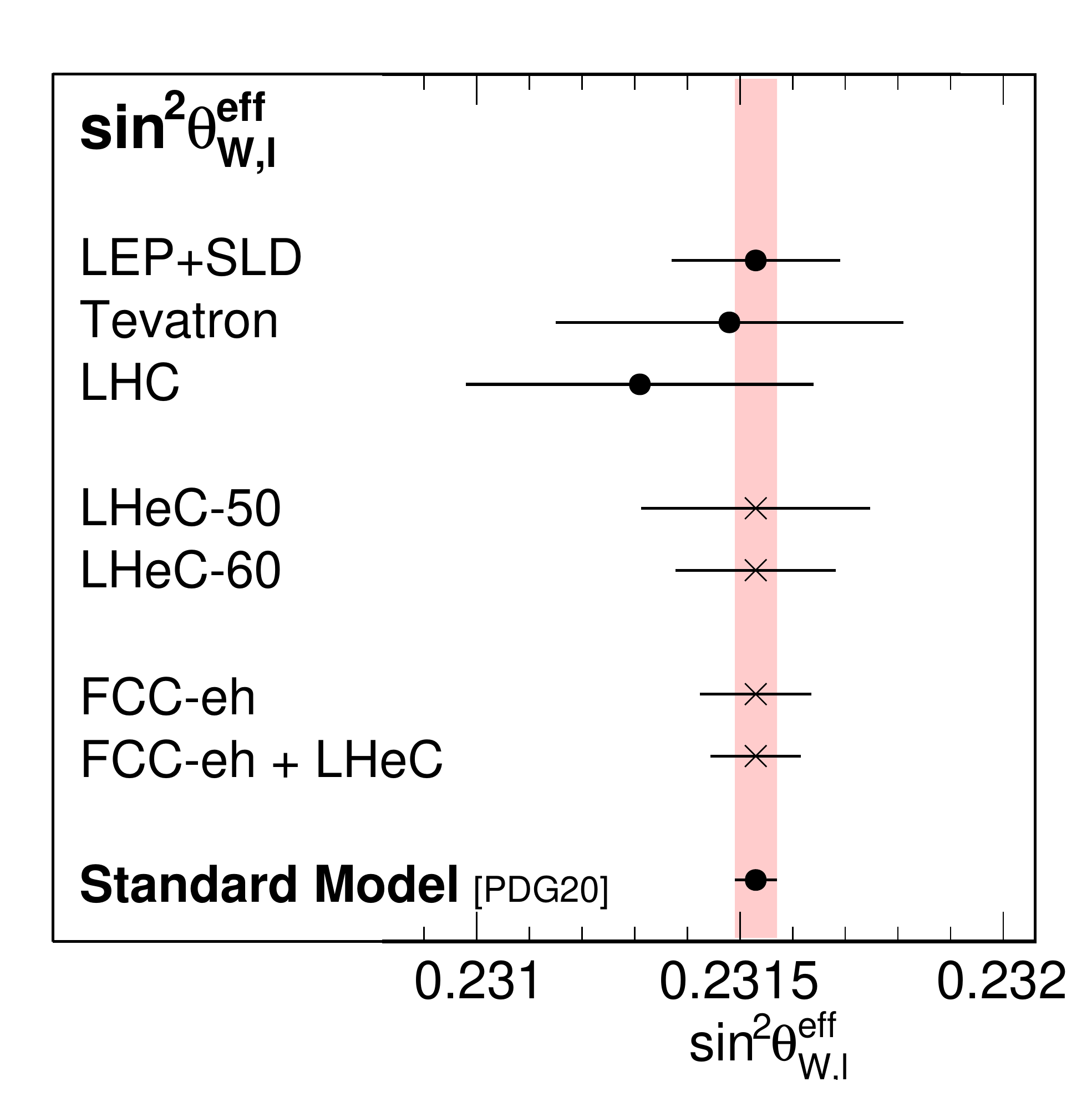}
    \caption{
      Prospected uncertainties for the determination of
      \sweffl\ from LHeC or FCC-eh in comparison with present
      measurements and the SM expectation.
    }
    \label{fig:mwsw2}
\end{figure}

A determination of \sweffl is further studied for LHeC data, and 
in a combined analysis of LHeC and FCC-eh
data. Results are presented in Tab.~\ref{tab:mw}.
\begin{table}[htb!]
  \centering
  \begin{tabular}{lcc}
    \toprule
    Facility   & $\Delta\sweffl$  & $\Delta\mW$    \\
    \midrule                      
    LHeC-50    &   $\pm0.00021$   & $\pm8\,\MeV$   \\
    LHeC-60    &   $\pm0.00015$   & $\pm6\,\MeV$   \\
    FCC-eh     &   $\pm0.00011$   & $\pm4.5\,\MeV$ \\
    FCC+LHeC   &   $\pm0.000086$  & $\pm3.6\,\MeV$ \\
    \bottomrule
  \end{tabular}
  \caption{
    Comparision of the prospected uncertainties on the determination
    of the effective weak mixing angle \sweffl\ for different
    collections of future DIS data.
    The right column shows the prospected uncertainties in \mW\ 
    when employing the on-shell renormalization scheme.
  }
  \label{tab:mw}
\end{table}
It is observed, that the uncertainties will be competitive with the
combination of the LEP+SLD data, but determined here from a single
experiment and a single process only.
At this level of precision, theoretical uncertainties may be of
non-negligible size, and should be assessed once real data are available.

It should be stressed that the determination of \sweffl\ represents
a precision test of the SM at the quantum level and
many BSM models predict modifications not only of \sweffl\ but also of the
form factors $\rho_{\text{NC}, f}$.
Unfortunately, in DIS, the contributions from the latter cannot be
removed by using asymmetries, like the  forward-backward asymmetry, as it can be
done for $Z$-pole measurements.
Therefore, a simultaneous determination of the
form factors $\rho_{\text{NC}, f}$ and \sweffl\ is of interest and it
is studied in terms of a
PDF+$\rho^\prime_{\text{NC}, f}$+$\kappa^\prime_f$ fit,
where the primed parameters represent multiplicative anomalous form
factors describing physics beyond the SM, i.e.\ they are unity in the
SM~\cite{Spiesberger:2018vki}.
The prospected uncertainties from the FCC-eh are compared with those from the LHeC and
with what was obtained from the combination of LEP and SLD data~\cite{ALEPH:2005ab}.
It is observed, that the level of precision which can be achieved in DIS is comparable with that from $e^+e^-$ experiments.  

\begin{figure}[tbh!]
    \centering
    \raisebox{-4mm}{
    \includegraphics[width=0.43\textwidth]{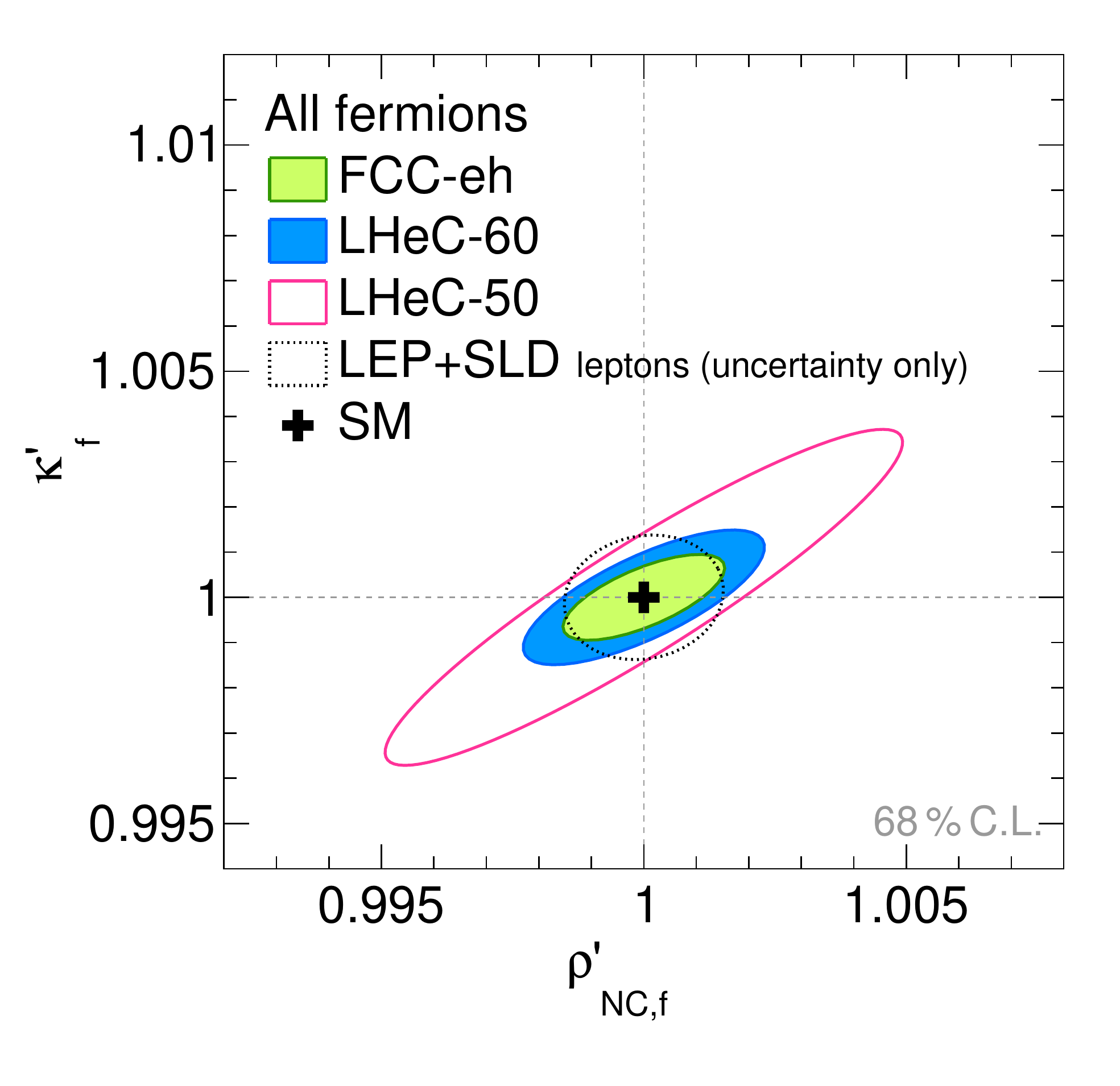}.pdf}
    \hspace{0.06\textwidth}
    \includegraphics[width=0.38\textwidth]{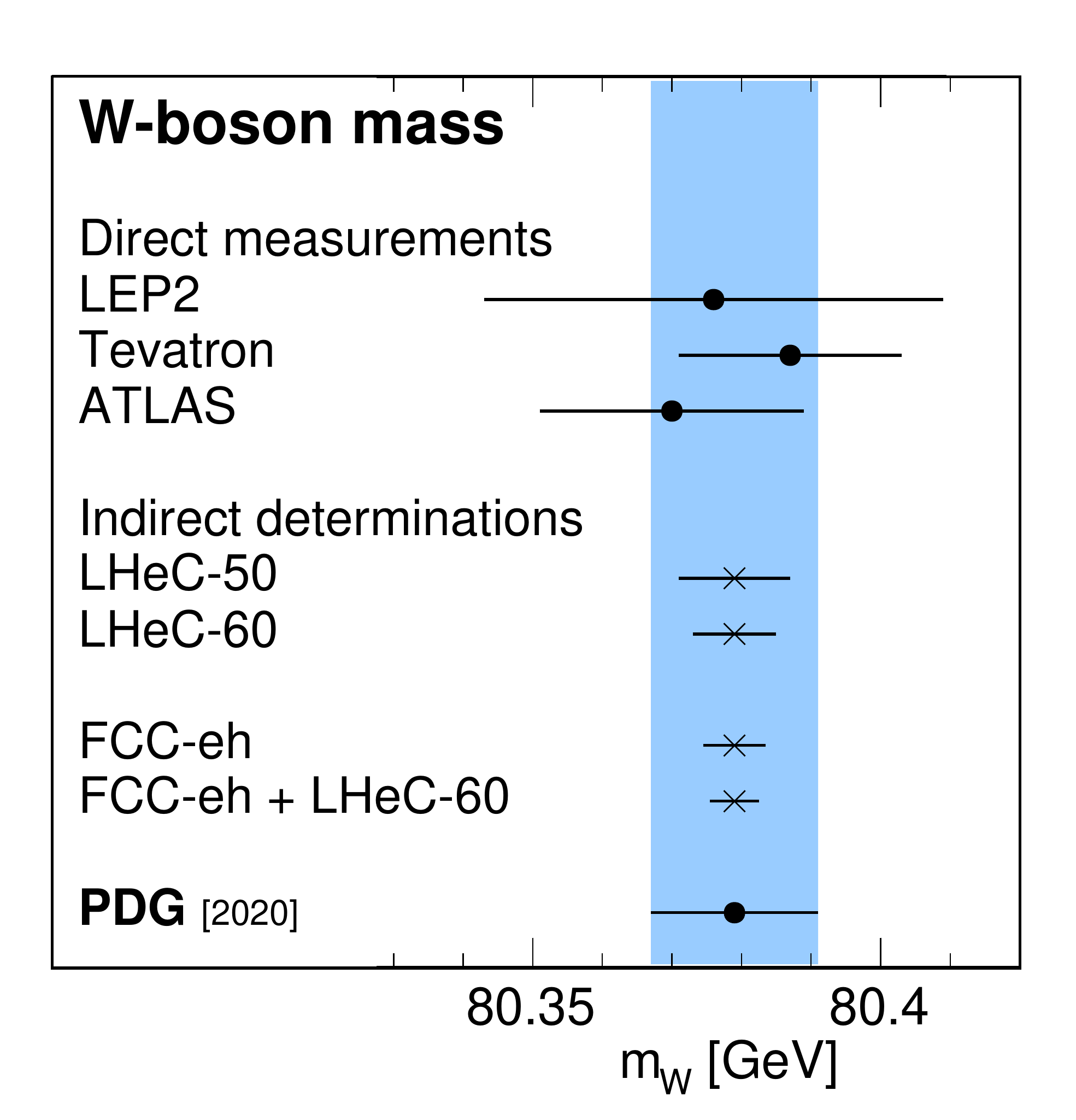}
    \caption{
      Left: Prospected uncertainties for the determination of
      $\rho_{\text{NC,}f}$ and \sweffl, expressed in terms of 
      anomalous form factors $\rho^\prime_{\text{NC,}f}$ and
      $\kappa^\prime_f$, in comparison with achieved uncertainties from 
      the LEP/SLD combination at a confidence level of 68\,\%. 
      Right: Prospected uncertainties for \mw\ when determined in the
      on-shell scheme in comparison with direct measurements.
    }
    \label{fig:mwsw2}
\end{figure}
%

\subsection{\boldmath The $W$-boson mass in the on-shell scheme}
In the on-shell renormalization scheme the $W$-boson mass is a
Born-level parameter together with $\alpha$ and $\mZ$, and
consequently any derived quantity, like e.g.\ \gf\ or \sw, 
depend on the value of \mw.
A PDF+\mW\ fit in the on-shell scheme is studied, which represents a
consistency test of electroweak theory at hand of a single
parameter only, when compared to direct measurements.
%
The prospected uncertainties in \mw\ are summarized in Tab.~\ref{tab:mw} and
displayed in Fig.~\ref{fig:mwsw2}.
We find, that at the FCC-eh, the value of \mw\ can be determined from the
inclusive DIS data with an uncertainty of 
\begin{equation}
  \Delta\mw = \pm\,4.5\,\MeV\,.
\end{equation}
The expectations for the LHeC are discussed in
Refs.~\cite{Britzger:2020kgg,LHeC:2020van}.
From a simultaneous analysis of LHeC and FCC-eh data, the prospected
 uncertainty is $\Delta\mw = \pm\,3.6\,\MeV$.
It is observed that the measurements at the LHeC or FCC-eh yield
smaller uncertainties than any direct measurement today, as well as
the indirect constraints from the global electroweak
fit~\cite{Haller:2018nnx,deBlas:2016ojx,Erler:2019hds}.
However, additional theoretical uncertaities will have to be taken
into account at that level of precision, and, for instance, the uncertainty due to the uncertainty of the top-quark
mass, $\Delta\mt=\pm0.5\,\GeV$, will contribute with an uncertainty of
about $\Delta\mW=2.5\,\MeV$ for the LHeC analysis.
Also the interpretation of the value of \mW\ is non-trivial.
In a dedicated study it was found that the
dominant contributions arise from the weak mixing angle in the NC
vector couplings, whereas the sensitivity from the $W$-boson
propagator in the CC DIS cross sections is comparatively small.

\subsection{\boldmath The oblique parameters $S$, $T$ and $U$}
\begin{figure}[tbh!]
    \centering
    \includegraphics[width=0.325\textwidth]{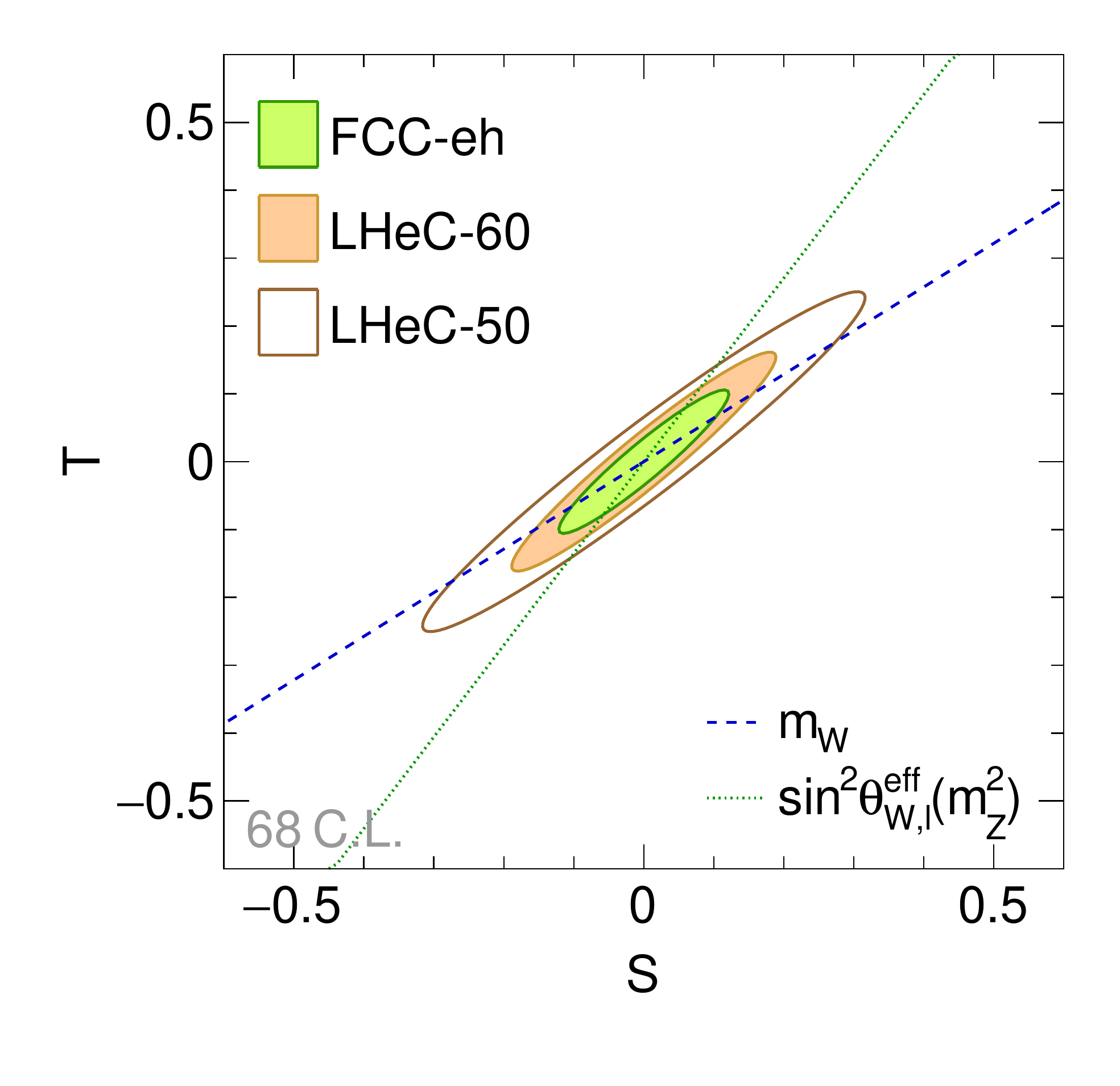}
    \includegraphics[width=0.325\textwidth]{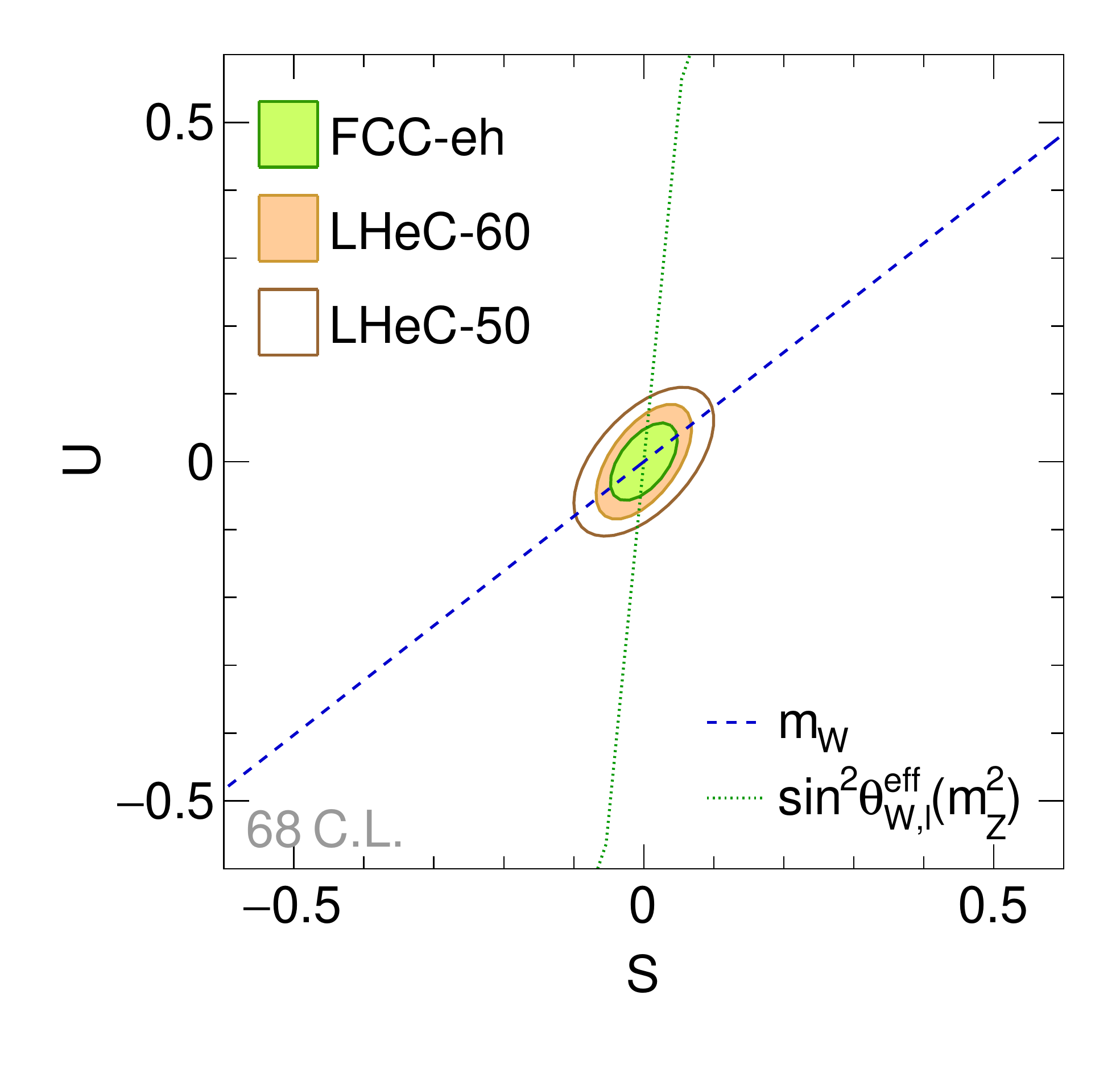}
    \includegraphics[width=0.325\textwidth]{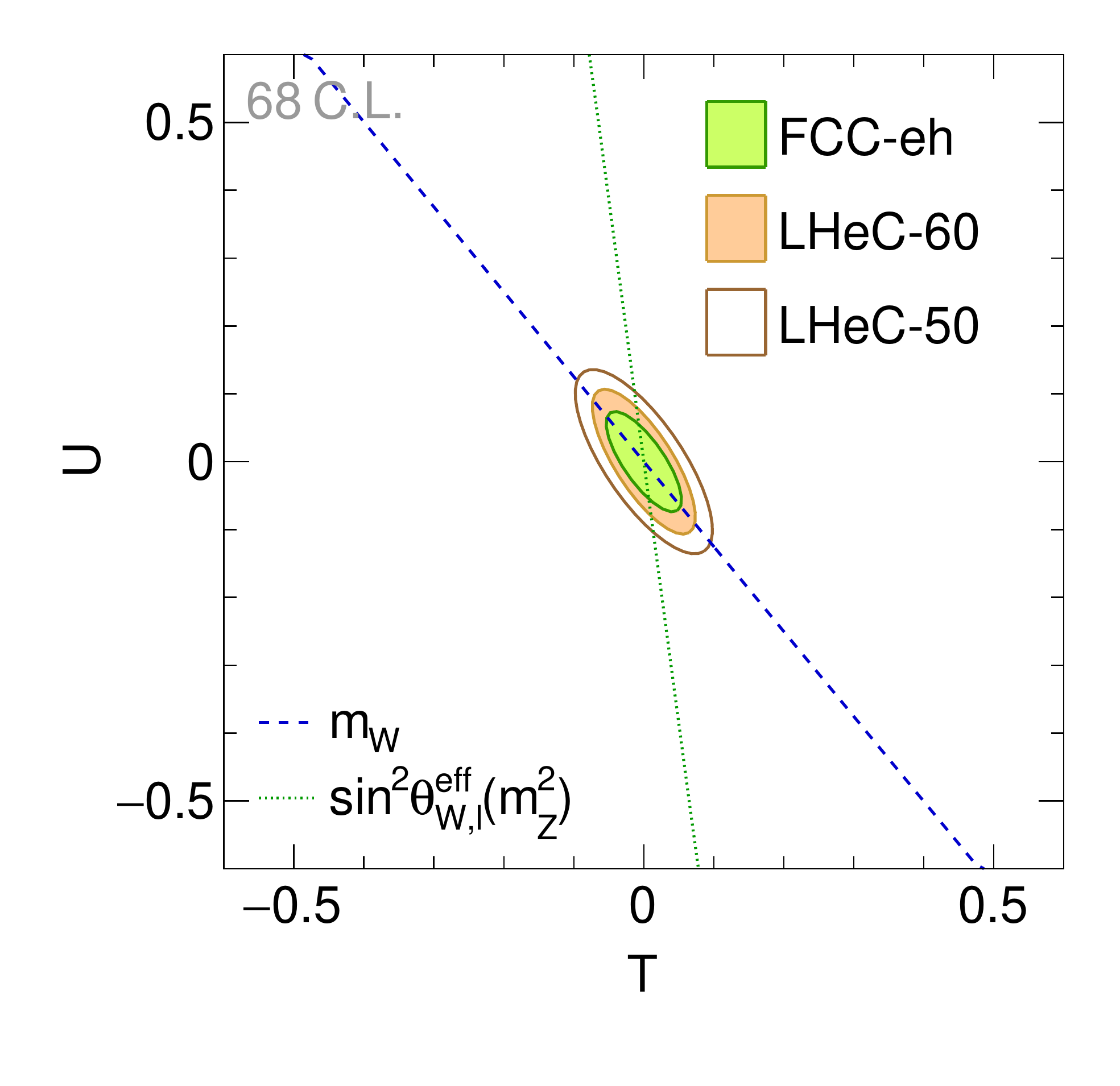}
    \caption{
      Prospected uncertainties on the oblique parameters $S$, $T$ and
      $U$ when two parameters are determined at a time (together with
      the PDFs) from LHeC or FCC-eh pseudo-data. The parameters are
      defined in the modified on-shell scheme and the lines indicate
      the SM values of \mw\ and \sweffl.
    }
    \label{fig:STU}
\end{figure}

Additional contributions from modifications of the SM can generically be considered in terms of the
oblique parameters $S$, $T$ and $U$~\cite{Peskin:1991sw}, which
represent non-SM contributions to the gauge boson self energies, and are 
particularly suitable to describe the effects due to new heavy particles.
We perform fits of two of these parameters together with parameters of
the PDFs.
The prospected uncertainties of two oblique parameters are displayed in
Fig.~\ref{fig:STU} at a confidence level of 68\,\% for LHeC and
FCC-eh simulated data.
The oblique parameters can be determined with uncertainties of about $\pm0.1$.
The combined analysis of NC and CC DIS data would also allow to determine
all three oblique parameters at a time, albeit with large
uncertainties due to correlations.

These studies were done in the modified on-shell scheme~\cite{Marciano:1980pb}, where \gf\ is
a Born-level quantity and \mw\ a prediction.
The on-shell renormalization scheme results in stronger correlations
of the oblique parameters.

\subsection{The running of the weak mixing angle}
The scale dependence of the weak mixing angle is given through the
renormalization scheme and loop corrections, and it is closely related
to the underlying gauge structure of electroweak theory. 
Many BSM models will yield a change of
the scale dependence of \sweff\ through modifications to the virtual corrections.
However, precision measurements are available only at
the $Z$-pole, or with moderate precision at much lower scales, which
is why the running of \sweff\ is largely not tested by experiments
until today.

The spacelike momentum transfer in DIS provides a unique opportunity
to perform a study of the scale-dependence of the weak mixing angle.
We perform a sensitivity study, by performing a single fit to
inclusive NC and CC DIS data and determining \sweff\ at different
values of \Qsq, together with the parameters of the PDFs.
Again, only the effective weak mixing angle in the vector-couplings are tested,
while other parameters like $\rho_\text{NC}$, $\rho_{CC}$ or $\Delta
r$ are taken from the SM.

\begin{figure}[tbh!]
    \centering
    \includegraphics[width=0.82\textwidth]{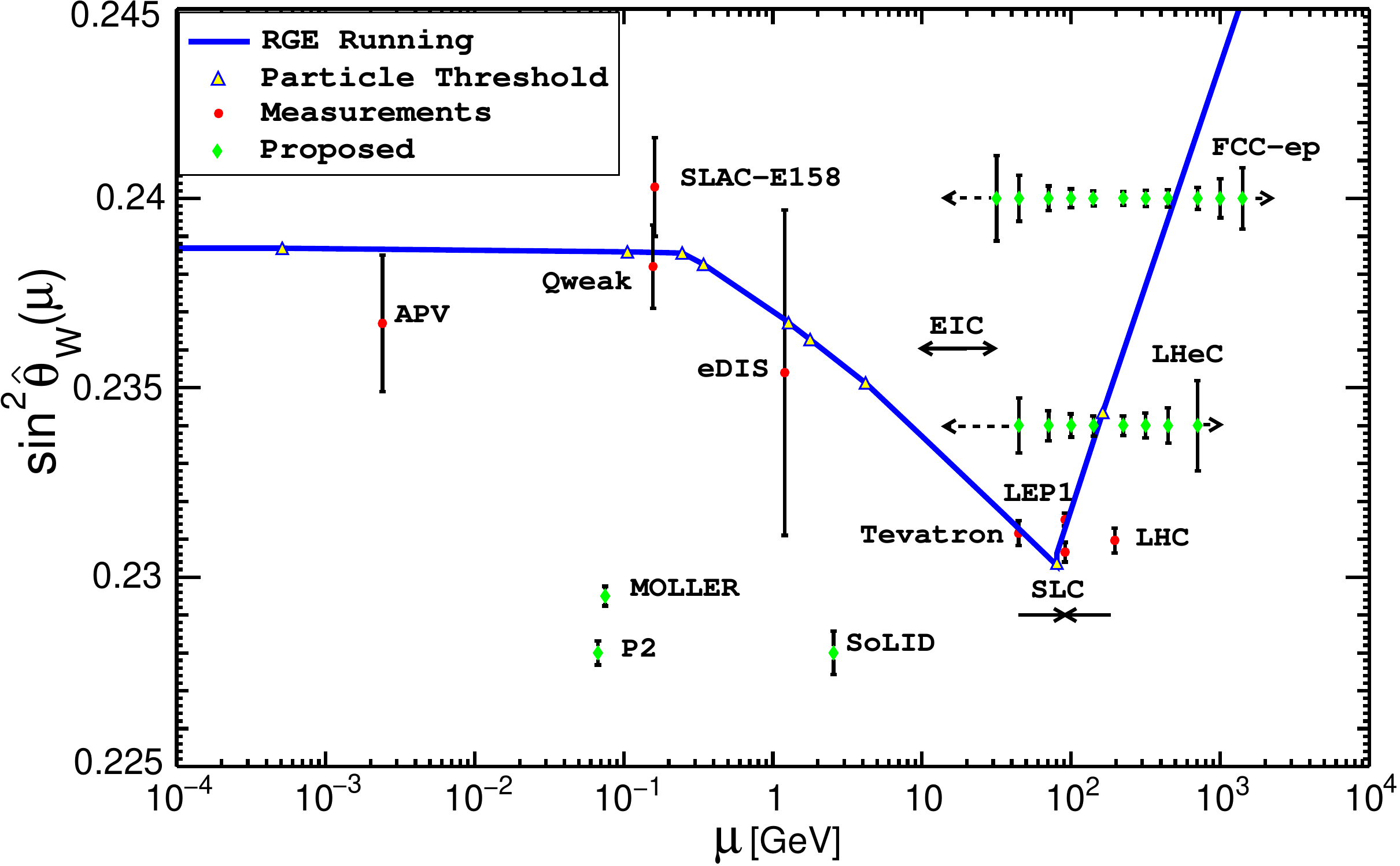}
    \caption{
      Present and future measurements of the running of the weak mixing angle in the
      $\overline{\text{MS}}$ scheme as a function of $\mu$.
      The prospected uncertainties from the LHeC and FCC-eh are
      displayed. All future facilities are displayed with green
      markers and have an arbitrary vertical offset for better
      visibility, whereas the $Z$-pole measurements are horizontally
      displaced. (Credits/courtesy: J.~Erler, R.~Ferro-Hernandez and X. Zheng; updated from Ref.~\cite{Erler:2004in,Erler:2017knj}).
    }
    \label{fig:sw2running}
\end{figure}
The expected uncertainties for LHeC and FCC-eh are displayed in
Fig.~\ref{fig:sw2running} and compared to available measurements and
the expectation in the $\overline{\text{MS}}$ scheme.
Small correction factors to translate the on-shell definition to the
$\overline{\text{MS}}$ definition of \sw\ will not visibly change the
uncertainty estimate.
It is found, that the LHeC and FCC-eh are able to measure the
running of the weak mixing angle in the range from about 30\,\GeV up
to the TeV scale with considerable precision.
Experiments at the LHeC and FCC-eh would represent the first measurements, where the
running is directly tested in a single experiment, and the first
indication for a scale dependence at the electroweak scale and above.

\subsection{Precision electroweak physics with charged currents}
Precision electroweak physics in charged-current interactions is a
poorly studied field, since measurements of
charged currents have commonly only low precision due to the presence of a 
neutrino that escapes undetected.
In contrast, in DIS, the kinematic variables \Qsq, $x$ and $y$ can be meaured 
from the incoming electron beam and the hadronic final state also for
charged-current DIS, which allows for precise measurements in a large 
kinematic region.

Similar to the NC case, EW corrections are included in the cross
section calculation through two form factors $\rho_{\text{CC},eq}$ and
$\rho_{\text{CC},e\bar{q}}$~\cite{Bohm:1987cg,Bardin:1989vz,Spiesberger:2018vki}.
We perform a study for their determinations, by introducing
corresponding multiplicative anomalous
form factors $\rho^\prime_{\text{CC},eq}$ and
$\rho^\prime_{\text{CC},e\bar{q}}$~\cite{Spiesberger:2018vki}, which 
 are unity in the SM.

\begin{figure}[tbh!]
    \centering
    \includegraphics[width=0.42\textwidth]{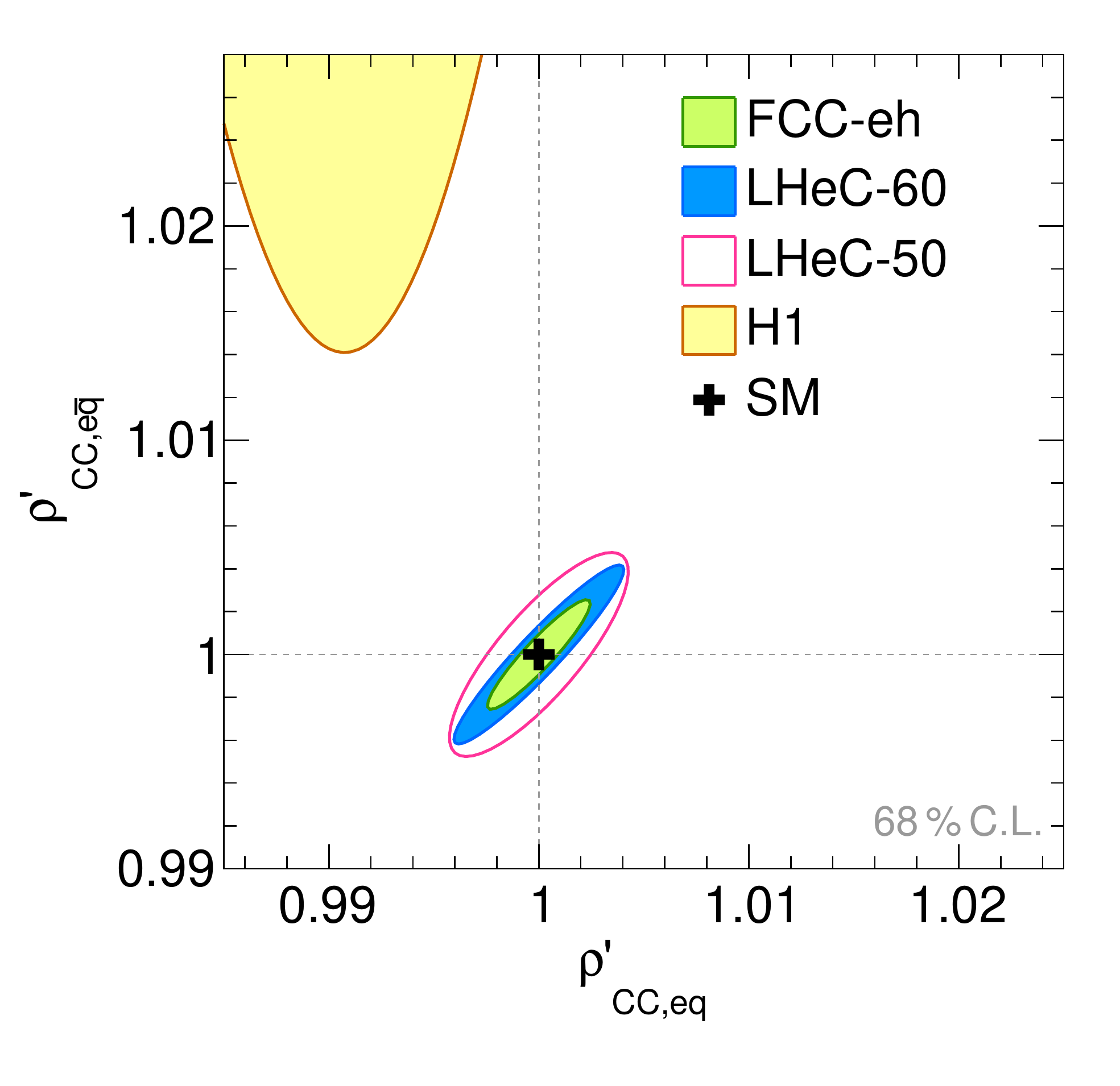}
    \includegraphics[width=0.42\textwidth]{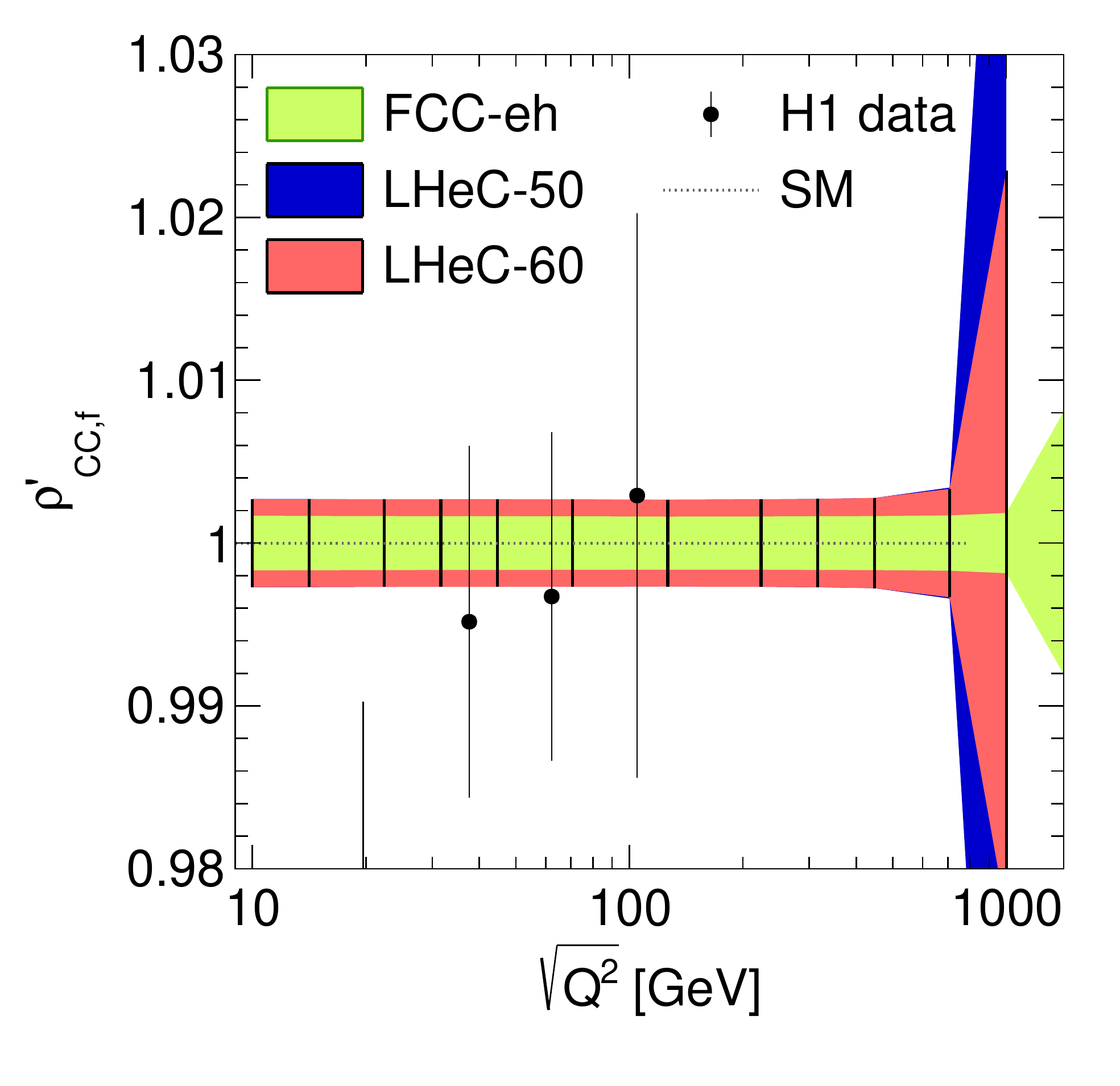}
    \caption{
      Left: Prospected uncertainties of CC coupling parameters
      $\rho^\prime_{\text{CC},eq}$ and
      $\rho^\prime_{\text{CC},e\bar{q}}$ from LHeC or FCC-eh data 
      compared with the only available measurement from
      H1~\cite{Spiesberger:2018vki}.
      Right: Prospected uncertainties for the scale dependent
      determination of $\rho^\prime_{\text{CC}}$ in comparison to H1.
    }
    \label{fig:CC}
\end{figure}
The prospected uncertainties of $\rho^\prime_{\text{CC},eq}$ and
$\rho^\prime_{\text{CC},e\bar{q}}$ are determined in a
PDF+$\rho^\prime_{\text{CC},eq}$+$\rho^\prime_{\text{CC},e\bar{q}}$
fit and are displayed in Fig.~\ref{fig:CC} (left) at a confidence
interval of 68\,\%.
It is observed that virtual corrections in CC interactions can be
tested at the per mille level, greatly exceeding the pioneering
measurements from H1~\cite{Spiesberger:2018vki}.

The $\rho_\text{CC}$ form factors are scale dependent, and it
is expected that BSM contributions will modify them through
contributions to these virtual corrections. 
The uncertainty estimates of $\rho_\text{CC}$ (using
$\rho^\prime_{\text{CC},eq}=\rho^\prime_{\text{CC},e\bar{q}}=\rho_\text{CC}$)
at different scales \Qsq\ is presented in Fig.~\ref{fig:CC} (right) and it
is seen that with FCC-eh uncertainties at the level of a few per mille are obtained
over a large range of $\sqrt{\Qsq}$, up to the TeV regime.
This is a very unique measurement.


\section{Conclusions and Outlook}
We have presented a study of the determination of parameters of the
Electroweak Theory using pseudo-data of the proposed LHeC and FCC-eh
electron-proton colliders.
The data are inclusive neutral-current and charged-current DIS cross sections
and include a full set of statistical and systematic uncertainties.
The study employs NNLO QCD predictions for the structure functions and
NLO EW predictions for the partonic cross sections. Parameters are
determined in a combined PDF plus EW fit.

It is found, that several quantities in the EW sector can be
determined with very high precision, and values of the effective weak
mixing angle or the $W$-boson mass may even exceed the precision of the
combined measurements from LEP/SLD. 
Since the interaction in DIS is mediated in the spacelike regime, 
all these measurements are unique.
In fact, several aspects of the EW theory can only be studied in DIS in the
$t$-channel.
This includes the light-quark weak-neutral-current
couplings, the running of the weak mixing angle from the electroweak 
up to the TeV energy scale, and precision electroweak physics in
charged-current interactions.


\footnotesize
\bibliography{britzger_epshep21}

\providecommand{\href}[2]{#2}\begingroup\raggedright\begin{thebibliography}{10}

\bibitem{AbelleiraFernandez:2012cc}
{LHeC Study Group}, J.~Abelleira~Fernandez {\em et~al.}, ``{A Large Hadron
  Electron Collider at CERN: Report on the Physics and Design Concepts for
  Machine and Detector},''
  \href{http://dx.doi.org/10.1088/0954-3899/39/7/075001}{{\em J. Phys. G} {39}
  (2012) 075001}, \href{http://arxiv.org/abs/1206.2913}{{\ttfamily
  arXiv:1206.2913}}.

\bibitem{LHeC:2020van}
{ LHeC, FCC-he Study Group} Collaboration, P.~Agostini {\em et~al.}, ``{The
  Large Hadron-Electron Collider at the HL-LHC},''
  \href{http://dx.doi.org/10.1088/1361-6471/abf3ba}{{\em J. Phys. G} {48}
  (2021) 110501}, \href{http://arxiv.org/abs/2007.14491}{{\ttfamily
  arXiv:2007.14491}}.

\bibitem{Abada:2019lih}
{FCC Study Group}, A.~Abada {\em et~al.}, ``{FCC Physics Opportunities}:
  {Future Circular Collider Conceptual Design Report Volume 1},''
  \href{http://dx.doi.org/10.1140/epjc/s10052-019-6904-3}{{\em Eur. Phys. J. C}
  {79} (2019) 474}.

\bibitem{Britzger:2020kgg}
D.~Britzger, M.~Klein and H.~Spiesberger, ``{Electroweak physics in inclusive
  deep inelastic scattering at the LHeC},''
  \href{http://dx.doi.org/10.1140/epjc/s10052-020-8367-y}{{\em Eur. Phys. J. C}
  {80} (2020) 831}, \href{http://arxiv.org/abs/2007.11799}{{\ttfamily
  arXiv:2007.11799}}.

\bibitem{Klein:1983vs}
M.~Klein and T.~Riemann, ``{Electroweak interactions probing the nucleon
  structure},''
\href{http://dx.doi.org/10.1007/BF01571719}{{\em Z. Phys.} {C24} (1984) 151}.

\bibitem{Sirlin:1980nh}
A.~Sirlin, ``{Radiative corrections in the SU(2)$_{\rm L} \times$ U(1) theory:
  A simple renormalization framework},''
\href{http://dx.doi.org/10.1103/PhysRevD.22.971}{{\em Phys. Rev.} {D22} (1980)
  971}.

\bibitem{Bohm:1986na}
M.~B{\"o}hm and H.~Spiesberger, ``{Radiative corrections to neutral current
  deep inelastic lepton nucleon scattering at HERA energies},''
\href{http://dx.doi.org/10.1016/0550-3213(87)90624-9}{{\em Nucl. Phys.} {B294}
  (1987) 1081}.

\bibitem{Bardin:1988by}
D.~{\relax Yu}. Bardin, C.~Burdik, P.~C. Khristova and T.~Riemann,
  ``{Electroweak radiative corrections to deep inelastic scattering at HERA,
  neutral current scattering},''
\href{http://dx.doi.org/10.1007/BF01557676}{{\em Z. Phys.} {C42} (1989) 679}.

\bibitem{Hollik:1992bz}
W.~Hollik, D.~{\relax Yu}. Bardin, J.~Bl{\"u}mlein, B.~A. Kniehl, T.~Riemann
  and H.~Spiesberger, ``{Electroweak parameters at HERA: Theoretical
  aspects},'' in {\em {Workshop on physics at HERA Hamburg, Germany, October
  29-30, 1991}},
1992.
\newblock

\bibitem{Blumlein:1990dj}
J.~Blümlein and M.~Klein, ``{Kinematics and resolution at future e p
  colliders},'' in {\em {1990 DPF Summer Study on High-energy Physics: Research
  Directions for the Decade (Snowmass 90)}}, 9 1990.

\bibitem{Yan:2021htf}
B.~Yan, Z.~Yu and C.~P. Yuan, ``{The anomalous Zbb\textasciimacron{} couplings
  at the HERA and EIC},''
  \href{http://dx.doi.org/10.1016/j.physletb.2021.136697}{{\em Phys. Lett. B}
  {822} (2021) 136697}, \href{http://arxiv.org/abs/2107.02134}{{\ttfamily
  arXiv:2107.02134}}.

\bibitem{Botje:2010ay}
M.~Botje, ``{QCDNUM: fast QCD evolution and convolution},''
  \href{http://dx.doi.org/10.1016/j.cpc.2010.10.020}{{\em Comput. Phys.
  Commun.} {182} (2011) 490}, \href{http://arxiv.org/abs/1005.1481}{{\ttfamily
  arXiv:1005.1481}}.
[Erratum: \href{http://arxiv.org/abs/1602.08383}{{\ttfamily
  arXiv:1602.08383}}].

\bibitem{Spiesberger:1995pr}
H.~Spiesberger, ``{EPRC: A program package for electroweak physics at HERA},''
  in {\em {Future physics at HERA. Proceedings, Workshop, Hamburg, Germany,
  September 25, 1995-May 31, 1996. Vol. 1, 2}},
1995.
\newblock

\bibitem{James:1975dr}
F.~James and M.~Roos, ``{Minuit: A system for function minimization and
  analysis of the parameter errors and correlations},''
\href{http://dx.doi.org/10.1016/0010-4655(75)90039-9}{{\em Comput. Phys.
  Commun.} {10} (1975) 343}.

\bibitem{ALEPH:2005ab}
{DELPHI, ALEPH, SLD, OPAL and L3 Collaborations and SLD and LEP Electroweak
  Working Groups}, S.~Schael {\em et~al.}, ``{Precision electroweak
  measurements on the $Z$ resonance},''
  \href{http://dx.doi.org/10.1016/j.physrep.2005.12.006}{{\em Phys. Rept.}
  {427} (2006) 257},
\href{http://arxiv.org/abs/hep-ex/0509008}{{\ttfamily arXiv:hep-ex/0509008}}.

\bibitem{Abazov:2011ws}
{ D0} Collaboration, V.~M. Abazov {\em et~al.}, ``{Measurement of
  $\sin^2\theta_{\rm eff}^{\ell}$ and $Z$-light quark couplings using the
  forward-backward charge asymmetry in $p\bar{p} \to Z/\gamma^{*} \to
  e^{+}e^{-}$ events with ${\cal L}=5.0$ fb$^{-1}$ at $\sqrt{s}=1.96$ TeV},''
  \href{http://dx.doi.org/10.1103/PhysRevD.84.012007}{{\em Phys. Rev.} {D84}
  (2011) 012007},
\href{http://arxiv.org/abs/1104.4590}{{\ttfamily arXiv:1104.4590}}.

\bibitem{Spiesberger:2018vki}
{ H1} Collaboration, V.~Andreev {\em et~al.}, ``{Determination of electroweak
  parameters in polarised deep-inelastic scattering at HERA},''
  \href{http://dx.doi.org/10.1140/epjc/s10052-018-6236-8}{{\em Eur. Phys. J.}
  {C78} (2018) 777},
\href{http://arxiv.org/abs/1806.01176}{{\ttfamily arXiv:1806.01176}}.

\bibitem{EW:PDG2020}
J.~Erler and A.~Freitas,
  \href{http://dx.doi.org/10.1093/ptep/ptaa104}{``{Electroweak Model and
  Constraints on New Physics},''} in {\em {Review of Particle Physics, PTEP
  2020 (2020) 8, 083C01}}, P.~A. Zyla {\em et~al.} (eds.), 2020.

\bibitem{Haller:2018nnx}
J.~Haller, A.~Hoecker, R.~Kogler, K.~Mönig, T.~Peiffer and J.~Stelzer,
  ``{Update of the global electroweak fit and constraints on two-Higgs-doublet
  models},'' \href{http://dx.doi.org/10.1140/epjc/s10052-018-6131-3}{{\em Eur.
  Phys. J.} {C78} (2018) 675},
\href{http://arxiv.org/abs/1803.01853}{{\ttfamily arXiv:1803.01853}}.

\bibitem{deBlas:2016ojx}
J.~de~Blas, M.~Ciuchini, E.~Franco, S.~Mishima, M.~Pierini, L.~Reina and
  L.~Silvestrini, ``{Electroweak precision observables and Higgs-boson signal
  strengths in the Standard Model and beyond: present and future},''
  \href{http://dx.doi.org/10.1007/JHEP12(2016)135}{{\em JHEP} {12} (2016) 135},
\href{http://arxiv.org/abs/1608.01509}{{\ttfamily arXiv:1608.01509}}.

\bibitem{Erler:2019hds}
J.~Erler and M.~Schott, ``{Electroweak Precision Tests of the Standard Model
  after the Discovery of the Higgs Boson},''
  \href{http://dx.doi.org/10.1016/j.ppnp.2019.02.007}{{\em Prog. Part. Nucl.
  Phys.} {106} (2019) 68}, \href{http://arxiv.org/abs/1902.05142}{{\ttfamily
  arXiv:1902.05142}}.

\bibitem{Peskin:1991sw}
M.~E. Peskin and T.~Takeuchi, ``{Estimation of oblique electroweak
  corrections},''
\href{http://dx.doi.org/10.1103/PhysRevD.46.381}{{\em Phys. Rev.} {D46} (1992)
  381}.

\bibitem{Marciano:1980pb}
W.~J. Marciano and A.~Sirlin, ``{Radiative corrections to neutrino induced
  neutral current phenomena in the SU(2)$_{\rm L} \times$ U(1) theory},''
  \href{http://dx.doi.org/10.1103/PhysRevD.31.213,
  10.1103/PhysRevD.22.2695}{{\em Phys. Rev.} {D22} (1980) 2695}.
[Erratum: \emph{Phys. Rev.} D31 (1985) 213].

\bibitem{Erler:2004in}
J.~Erler and M.~J. Ramsey-Musolf, ``{The Weak mixing angle at low energies},''
  \href{http://dx.doi.org/10.1103/PhysRevD.72.073003}{{\em Phys. Rev. D} {72}
  (2005) 073003}, \href{http://arxiv.org/abs/hep-ph/0409169}{{\ttfamily
  arXiv:hep-ph/0409169}}.

\bibitem{Erler:2017knj}
J.~Erler and R.~Ferro-Hern\'andez, ``{Weak Mixing Angle in the Thomson
  Limit},'' \href{http://dx.doi.org/10.1007/JHEP03(2018)196}{{\em JHEP} {03}
  (2018) 196}, \href{http://arxiv.org/abs/1712.09146}{{\ttfamily
  arXiv:1712.09146}}.

\bibitem{Bohm:1987cg}
M.~B{\"o}hm and H.~Spiesberger, ``{Radiative corrections to charged current
  deep inelastic electron - proton scattering at HERA},''
\href{http://dx.doi.org/10.1016/0550-3213(88)90652-9}{{\em Nucl. Phys.} {B304}
  (1988) 749}.

\bibitem{Bardin:1989vz}
D.~{\relax Yu}. Bardin, K.~C. Burdik, P.~K. Khristova and T.~Riemann,
  ``{Electroweak radiative corrections to deep inelastic scattering at HERA,
  charged current scattering},''
\href{http://dx.doi.org/10.1007/BF01548593}{{\em Z. Phys.} {C44} (1989) 149}.

\end{thebibliography}\endgroup

\end{document}